\documentstyle[12pt]{article}
\input epsf

\newcommand{\sect}[1]{\setcounter{equation}{0}\section{#1}}

\newcommand\M{{\cal M}}
\newcommand\N{{\cal N}}

\newcommand{\ssj}[6]{\left| \begin{array}{ccc} #1 & #2 & #3 \\ #4 & #5 & #6 \end{array} \right|} 

\newcommand{\f}[1]{ w_{#1_1}w_{#1_2}w_{#1_3}w_{#1_4}w_{#1_5}w_{#1_6}}

\def\tv{Turaev-Viro}

\def\inv{invariant}

\begin{document}

\begin{titlepage}

\begin{flushright}
DAMTP/96-94\\
November 1996\\
\end{flushright}

\vspace{.5in}

\begin{center}
{\LARGE \bf
 Building Blocks in Turaev-Viro Theory}\\

\vspace{.4in}
{\large {Radu Ionicioiu}
        \footnote{\it on leave from Institute of Gravitation and
        Space Sciences, 21-25 Mendeleev Street, 70168 Bucharest, Romania}\\
       {\small\it DAMTP}\\
       {\small\it University of Cambridge}\\
       {\small\it Silver Street, Cambridge, CB3 9EW, UK}\\
       {\small\it email: ri10001@damtp.cam.ac.uk}\\}
\end{center}

\vspace{.5in}
\begin{center}
{\large\bf Abstract}
\end{center}

\begin{center}
\begin{minipage}{4.75in}
{\small
We study the form of the Turaev-Viro partition function $Z(\M)$ for different \mbox{3-manifolds} with boundary. We show that for $S^2$ boundaries $Z(\M)$ factorizes into a term which contains the boundary dependence and another which depends only on the topology of the underlying manifold. From this follows easily the formula for the connected sum of two manifolds $Z(\M \# \N)$. For general $T_g$ boundaries this factorization holds only in a particular case.}
\end{minipage}
\end{center}

\end{titlepage}
\addtocounter{footnote}{-1}

\sect{Introduction \label{Intro}}

One of the most active fields in theoretical physics is quantum gravity. A good overview of the problems faced by a quantum theory of gravity and of different approaches to this subject is given in \cite{ish1}, \cite{ish2}.
Besides technical problems, such a theory is also faced with philosophical questions about our understanding of fundamental concepts like the nature of space, time and the problem of causality. In the same framework, we would like a final theory of gravity to be able to answer questions like 'Why is the dimension of space-time four?' or 'How did the Universe look at the Planck time?'.

From the many different approaches in this field (which include Euclidean quantum gravity, superstring theory, Ashtekar's program) we are interested here in the Turaev-Viro theory. This topic is closely related to the Atiyah-Witten approach of topological quantum field theories (TQFTs), knot theory, invariants of 3-manifolds, spin networks and other similar fields (for a link between quantum gravity and TQFT see \cite{jb}).

Ponzano and Regge \cite{pr} noticed a striking similarity between the asymptotic (large spin) value of the $6j$-symbols and the path integral (partition function) for three-dimensional quantum gravity. However, the Ponzano-Regge partition function for simplicial 3-d gravity is not finite. The regularized version of it is exactly the Turaev-Viro invariant.

In order to establish notation we define in what follows the Turaev-Viro invariant for a 3-manifold $\M$.
Consider a commutative ring $K$ with unity and denote by $K^*$ the set of its invertible elements. Assume we have a finite set $I$ (of 'colors'), a function $I \rightarrow K^*$ which maps every 'colour' $i$ to an element of $K^*$, $i \mapsto w_i \in K^*$, and a distinguished element $w$ of $K^*$. We also suppose that there is a set $adm$ of unordered triples $(i,j,k)$ of $I$ called $admissible$. The ordered 6-tuple $(i,j,k,l,m,n)$ is called admissible, if the unordered triples $(i,j,k),\ (k,l,m),\ (m,n,i)$ and $(j,l,n)$ are admissible. We associate to each admissible 6-tuple an element of $K$ called the $6j$-symbol and denoted by 

\[ \ssj{i}{j}{k}{l}{m}{n}\]

We assume that the $6j$-symbols have the following properties:\\

1) Symmetry:
 
\begin{equation}
\ssj{i}{j}{k}{l}{m}{n}=\ssj{j}{i}{k}{m}{l}{n}=\ssj{i}{k}{j}{l}{n}{m}=\ssj{i}{m}{n}{l}{j}{k}
\label{sym}
\end{equation}
\\

2) Orthogonality:

\begin{equation}
\sum_{k\in I} w^2_{k}\ w^2_{n} \ssj{i}{j}{k}{l}{m}{n} \ssj{i}{j}{k}{l}{m}{n'} = \delta_{n,n'}
\label{orto}
\end{equation}
\\
for all admissible 6-tuples $(i,j,k,l,m,n)$ and $(i,j,k,l,m,n')$.\\

3) They satisfy the Biedenharn-Elliot identity:

\begin{equation}
\sum_{l} w^2_{l} \ssj{j_1}{k_2}{k_3}{l}{m_3}{m_2} \ssj{j_2}{k_3}{k_1}{l}{m_1}{m_3} \ssj{j_3}{k_1}{k_2}{l}{m_2}{m_1} =
\ssj{j_1}{j_2}{j_3}{k_1}{k_2}{k_3} \ssj{j_1}{j_2}{j_3}{m_1}{m_2}{m_3}
\label{be}
\end{equation}

The initial data also satisfy the following identities:\\

4) for any $i \in I$

\begin{equation}
w^2\ w^2_i= \sum_{j,k:\ (i,j,k)\in adm} w^2_j\ w^2_k \equiv \sum_{j,k} w^2_j\ w^2_k\ \delta(i,j,k)
\label{ww}
\end{equation}
\\
where we used the notation of \cite{kms} and put $\delta(i,j,k)=1$ if $(i,j,k)\in adm$ and zero otherwise.
For irreducible initial data it is sufficient to verify the condition (\ref{ww}) only for one particular $i$. (The initial data are called {\it irreducible} if for any $i, j \in I$ there is a sequence $l_1,l_2,\ldots,l_n$ with $l_1=i,\ l_n=j$ and $\delta (l_k,l_{k+1},l_{k+2})=1$ for any $k=1,\ldots,n-2$). In what follows we assume the initial data are irreducible.
Denote by

\begin{equation}
\tilde w^2=\sum_{i \in I} w_i^4
\end{equation}

In \cite{kms} it was proved that the following identity holds

\begin{equation}
\sum_{i \in I} w_i^2\ \delta(i,j,k)= c\cdot w_j^2\ w_k^2
\end{equation}
where

\[ c \equiv \frac{w^2}{\tilde w^2}\]

In the particular case of quantum $6j$-symbols associated to $U_q(sl(2,C))$ we have $\tilde w^2=w^2$ and therefore we recover this case by setting $c=1$ in all formulae.

For any triangulation of $\M$, we can define a $colouring$ by the map

\[ \varphi:\{ E_1,E_2,\ldots, E_b \} \rightarrow I \]
\\
from the set of edges to the set of colors. To each coloured tetrahedron $t$ (by the map $\varphi$) we associate a $6j$-symbol denoted by $|T^{\varphi}_t| \in K$.
The \tv\ \inv\ (partition function) for the manifold $\M$ with the boundary coloring (denoted by $j$) kept fixed is:

\begin{equation}
Z(\M,j)= \sum_{\varphi\in adm(\M, \partial \M)} w^{-2a+e}\ \prod_{r=1}^f w_{\varphi(E_r)} \prod_{s=f+1}^b w^2_{\varphi(E_s)}\ \prod_{t=1}^d |T_t^{\varphi}|  
\label{ztv}
\end{equation}
where:\\
$a$ -- number of vertices of $\M$, from which $e$ of them belong to $\partial \M$ \\
$b$ -- number of edges of $\M$ and the first $f$ are on $\partial \M$ \\
$d$ -- number of tetrahedra of $\M$ \\

For a closed manifold ($\partial \M=\emptyset$) we simply put $e=f=0$ in eq.~(\ref{ztv}) and sum over all admissible colorings $\varphi \in adm(\M)$.

\sect{Surgeries on $\M$}

As seen in the last section, the \tv\ \inv\  for a manifold depends only on the topology of $\M$ and on the colouring of the boundary (if any). For a manifold $\M$ with boundaries $b_1,..,b_n$, we denote the corresponding invariant by $Z({\cal M};b_1,..,b_n)$.

The general formula (\ref{ztv}) is hard to work with, especially for large triangulations. In \cite{tv} the invariants for particular manifolds were derived using an alternative way. This was done by passing to the dual triangulation and working with simple 2-skeletons of 3 manifolds.

Our approach here is different. In order to find the \tv\ \inv\ $Z(\M)$ for a general $\M$, we try to find a set of simple (or 'elementary') manifolds from which any $\M$ can be constructed. Next we need a list of operations (e.g. adding/removing a handle, connected sum, gluing, making a boundary etc.) which can be applied to the set of elementary manifolds to yield an arbitrary $\M$. What we need then is the modification of $Z(\M)$ corresponding to such a surgery. 

We start by computing this invariant for some simple manifolds. The simplest case is the 3-sphere $S^3$, for which we have \cite{kms}:

\begin{equation}
Z(S^3)=\frac{1}{c\ w^2}
\label{zs3}
\end{equation}

Next we analyse the way $Z(\M)$ transforms under simple surgeries on $\M$.

\subsection{Making $S^2$ boundaries: the case of $S^3$}

We want now to calculate $Z$ for a 3-sphere $S^3$ with an arbitrary number of boundaries. The possible boundaries of a 3-manifold are closed 2-manifolds and their classification is complete. If we consider only orientable boundaries, these are uniquely specified by the genus $g$ (or, equivalently, by the Euler characteristic $\chi=2(1-g)$). Thus, the most general orientable boundary is a \mbox{g-genus} torus $T_g$ (with the natural convention that $T_0\equiv S^2$ and $T_1$ is the usual torus $T^2\equiv S^1 \times S^1$) or a disjoint sum of such components. In what follows, we restrict ourselves only to the case of $S^2$ boundaries.

For a manifold with boundary, $Z$ depends on the specific triangulation of $\partial \M$ and thus is a function of the number of boundary vertices.

We start with a few examples. Consider first an $S^2$ boundary triangulated with $N_0=4,\ N_1=6$ and $N_2=4$ (the surface of a tetrahedron; $N_k$ is the number of k-dimensional simplices). We will denote this by the (4,\ 6,\ 4) triangulation of $S^2$.

Straightforward calculations give the following result (see Appendix 1 for detailed calculations):

\begin{equation}
 Z(S^3;S^2_{(4,6,4)},j)=w^{-4}\ \f{j} \ssj{j_1}{j_2}{j_3}{j_6}{j_4}{j_5}
\label{zs31}
\end{equation}
\\
where $j\equiv (j_1,\ldots ,j_6)$ is a particular colouring of the boundary and the subscript $(4,6,4)$ represents the specific triangulation of $S^2$ (which will be omitted sometimes for simplicity).

The next step is to make another $S^2$ boundary, the resulting manifold being a hyper-cylinder ${\cal M} =S^2 \times [0,1]$. The partition function is:

\begin{eqnarray}
 Z(S^3;S^2,j,\ S^2,k)=c\ w^{-6}\ \f{j} \ssj{j_1}{j_2}{j_3}{j_6}{j_4}{j_5} \cdot \nonumber \\ \f{k} \ssj{k_1}{k_2}{k_3}{k_6}{k_4}{k_5}
\label{zs32}
\end{eqnarray}
\\
($j$ and $k$ are, respectively, the colourings of the two boundaries).

Comparing eqs.~(\ref{zs3}) - (\ref{zs32}) it is easy to see that adding an $S^2$ boundary (in the (4,\ 6,\ 4) triangulation) amounts to multiplying the partition function of $S^3$ by

\begin{equation}
 f(S^2_{(4,6,4)},j)=c\ w^{-2}\ \f{j} \ssj{j_1}{j_2}{j_3}{j_6}{j_4}{j_5}
\label{f}
\end{equation}

We call this the boundary function.

The proof of this property will be given in the next section.

We are now interested in the dependence of the boundary function $f$ on different triangulations $(N_0,\ N_1,\ N_2)$ of $S^2$. Not all of the $N_k$'s are independent, since they obey the Dehn-Sommerville relations (which give the conditions for a triangulation to be a closed simplicial manifold). In 2 dimensions, this is simply $2N_1=3N_2$. We have also another constraint, which fixes the topology to be a 2-sphere $\chi\equiv N_0-N_1+N_2=2$. Thus, all the triangulations of $S^2$ are of the form 

\[ (N_0,\ N_1,\ N_2) = (N_0,\ 3N_0-6,\ 2N_0-4) \]
\\
and the only free parameter is the number of vertices $N_0$.

The simplest way of constructing a 3-sphere with an $S^2$ boundary is by taking the cone over the boundary. For an $(N_0,\ N_1,\ N_2)$ triangulation of $S^2$, the corresponding triangulation will be $(N_0+1,\ N_1+N_0,\ N_2+N_1,\ N_2)$, or taking into account the last relation, this gives $(N_0+1,\ 4N_0-6,\ 5N_0-10,\ 2N_0-4)$. However, this construction does not work for a general $T_g,\ g>0$ boundary, since a cone over a boundary is a simplicial 3-manifold only if the boundary is $S^2$. 

Consider now two other examples, the (3,\ 3,\ 2) and the (5,\ 9,\ 6) triangulation of $S^2$. From Fig. \ref{s3s2} is easy to compute the corresponding \tv\ invariant (see Appendix 2 for details).

\begin{figure}
\epsffile{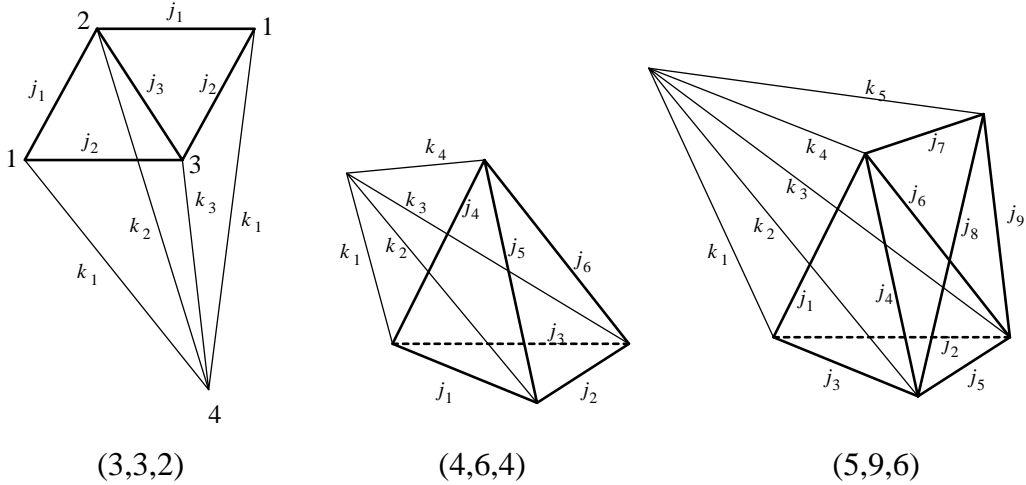}
\caption{Triangulation of a 3-ball ${\cal B}^3$ with different $(N_0,N_1,N_2)$ of the $S^2$ boundary (in bold).}
\label{s3s2}
\end{figure}

To summarize, we have obtained the following expressions for the boundary functions:

\[ f(S^2_{(3,3,2)},j)= c\ w^{-1} w_{j_1} w_{j_2} w_{j_3} \]
\[ f(S^2_{(4,6,4)},j)= c\ w^{-2} w_{j_1}\ldots w_{j_6} \ssj{j_1}{j_2}{j_3}{j_6}{j_4}{j_5} \]

\begin{equation}
f(S^2_{(5,9,6)},j)= c\ w^{-3} w_{j_1}\ldots w_{j_9} \ssj {j_4}{j_5}{j_6}{j_2}{j_1}{j_3} \ssj{j_4}{j_5}{j_6}{j_9}{j_7}{j_8}
\label{f123}
\end{equation}

We can derive now an important identity satisfied by the $f$ function. Suppose we glue together two 3-balls ${\cal B}^3$ (i.e. an $S^3$ with one $S^2$ boundary) along the common $S^2$ boundary. The resulting manifold will be a 3-sphere $S^3$. If the $S^2$ boundary is arbitrarily triangulated by $(N_0, N_1, N_2)$, we have

\[ Z(S^3)= \sum_{i}Z^2(S^3;S^2_{(N_0,N_1,N_2)},i) = Z^2(S^3) \sum_{i}f^2(S^2_{(N_0,N_1,N_2)},i) \]
\\
where we define

\[ f(S^2_{(N_0,N_1,N_2)},j)\equiv \frac{Z(S^3;S^2_{(N_0,N_1,N_2)},j)}{Z(S^3)} \]
\\
which thus satisfies:

\begin{equation}
\sum_{j}f^2(S^2_{(N_0,N_1,N_2)},j)=\frac{1}{Z(S^3)}= c\ w^2
\label{idf}
\end{equation}

It is easy to check that eq.~(\ref{idf}) is satisfied by all three boundary functions~(\ref{f123}).
An immediate result which follows from eqs.~(\ref{zs32}) and (\ref{f}) is the partition function for a three handle $S^2\times S^1$. By gluing together the two boundaries of the cylinder $S^2\times [0,1]$, we obtain:

\[ Z(S^2\times S^1)=\sum_{j}Z(S^3;S^2_{(4,6,4)},j,\ S^2_{(4,6,4)},j)= \]
\[ =Z(S^3) \sum_{j}f^2(S^2_{(4,6,4)},j) = 1 \]

Thus, we have:\\
\\
{\bf Corollary}

\begin{equation}
Z(S^2\times S^1)=1
\label{ztor}
\end{equation}

\subsection{Making $S^2$ boundaries: the case of arbitrary $\M$ \label{s2bound}}
 
Consider an arbitrary 3-manifold $\M$. Starting from $\M$ we want to compute the \tv\ \inv\ for a new manifold $\cal M'$ which is constructed from $\M$ by making an $S^2$ boundary. We want to show that we can factor out the contribution of the boundary.

{\bf Lemma}

Let $\M$ be an arbitrary 3-manifold and $Z(\M)$ the corresponding \tv\ invariant. The manifold $\M'$ which is constructed from $\M$ by cutting out an $S^2$ boundary in the (4,\ 6,\ 4) triangulation has the partition function

\begin{equation}
Z(\M')\equiv Z(\M;S^2_{(4,6,4)},j)=Z(\M)\cdot f(S^2_{(4,6,4)},j)
\label{s2factor}
\end{equation}
\\
where $f(S^2_{(4,6,4)},j)$ is given by~(\ref{f}) and depends only on the colouring $j$ of the $S^2$ boundary.
\\
\\
{\bf Proof:}

Let $\M$ be an arbitrary manifold with boundaries and let $Z(\M)$ be its partition function given by eq.~(\ref{ztv}). Denote by $\M'$ the manifold which is obtained from $\M$ by making an $S^2$ boundary in the $(4,6,4)$ triangulation.

\begin{figure}
\epsffile{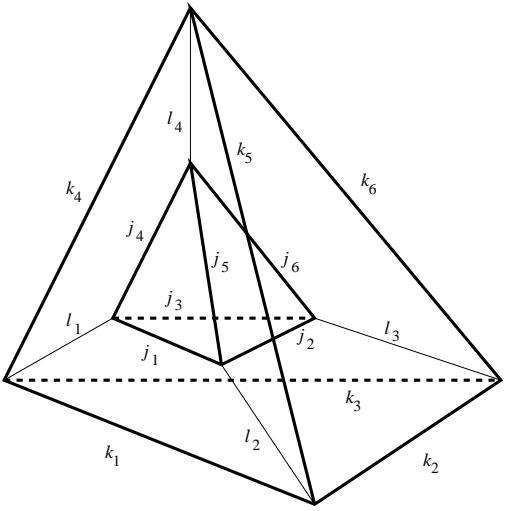}
\caption{Triangulation of the hypercylinder $S^2\times [0,1]$, viewed as a 3-sphere $S^3$ with two $S^2$ boundaries (in bold).}
\label{tetra1}
\end{figure}

Consider one of the tetrahedra, with edges labeled by $(k_1,\ldots,k_6)$, belonging to the interior of $\M$ (thus $Z(\M)$ includes a sum over $k_1,\ldots,k_6$). We replace this tetrahedron by the hypercylinder $S^2\times S^1$ given in Fig.~\ref{tetra1} and we obtain a triangulation for $\M'$. Each triangular prism corresponding to the four faces of the two $S^2$ boundaries in Fig.~\ref{tetra1} is further decomposed into three tetrahedra (see Fig.~\ref{tetra4}). With $a,b,d,e$ and $f$ defined for $\M$ as in~(\ref{ztv}), we have for the new manifold $\M'$:
\\
\\
$a'=a+4$,\ \ $e'=e+4$ (the number of vertices is increased by 4 and all of them are on the boundary $\partial \M'$); \\
$b'=b+16$ edges, $f'=f+6$; there are 16 new edges ($l_1,\ldots ,l_4$, $j_1, \ldots, j_6$, $m_1,\ldots,m_6$) from which six ($j_1,\ldots, j_6$) are on $\partial \M'$; \\
$d'=d+12-1$ tetrahedra (exclude 1 of them ($k_1,\ldots, k_6$) which was already in $\M$).

Thus, the partition function of $\M'$ is:

\[ Z(\M') = \sum_{\varphi\in adm(\M)} \sum_{\scriptsize \begin{array}{c}l_1...l_4 \\ m_1...m_6 \end{array}} w^{-2a+e}\ w^{-8+4}\prod_{r=1}^f w_{\varphi(E_r)}\ w_{j_1}\ldots w_{j_6} \cdot \]
\[ \prod_{s=f+1}^{b-6}w^2_{\varphi(E_s)}\ w^2_{k_1}\ldots w^2_{k_6}\ w^2_{l_1}\ldots w^2_{l_4}\ w^2_{m_1}\ldots w^2_{m_6} \cdot\]
\[  \prod_{t=1}^{d-1} |T_t^{\varphi}| \cdot \left( \begin {array}{c} product\ of\ twelve\ 6j-symbols \\ from\ the\ cylinder\ S^2\times I \end{array} \right) =\]

\[ = w^4 \sum_{\varphi\in adm(\M)} w^{-2a+e}\ \prod_{r=1}^f w_{\varphi(E_r)}\ 
\prod_{s=f+1}^{b-6}w^2_{\varphi(E_s)}\ w_{k_1}\ldots w_{k_6}\ \prod_{t=1}^{d-1} |T_t^{\varphi}| \cdot \]
\[ w^{-8} \sum_{\scriptsize \begin{array}{c}l_1...l_4 \\ m_1...m_6 \end{array}} w_{j_1}\ldots w_{j_6}\ w_{k_1}\ldots w_{k_6}\ w^2_{l_1}\ldots w^2_{l_4}\ \cdot \]
\[ w^2_{m_1}\ldots w^2_{m_6}\cdot \left( \begin {array}{c} product\ of\ twelve\ 6j-symbols \\ from\ the\ cylinder\ S^2\times I \end{array} \right) \]
\\

In the last formula we have explicitly factored out the terms which are involved in the partition function of the the hypercylinder $S^2\times S^1$ (see eq.~(\ref{big12j})). Thus, in the product over the internal edges of $\M$ the last six are those of the thetrahedron $(k_1,\ldots, k_6)$. Summing over $l_1,\ldots,l_4$ and $m_1,\ldots, l_4$ we get exactly the partition function~(\ref{zcil})

\[ Z(\M')= c\ w^{-2}\ w_{j_1}\ldots w_{j_6} \ssj{j_1}{j_2}{j_3}{j_6}{j_4}{j_5}
\sum_{\varphi\in adm(\M)} w^{-2a+e}\ \prod_{r=1}^f w_{\varphi(E_r)} \cdot \]
\[ \prod_{s=f+1}^{b-6}w^2_{\varphi(E_s)}\ w^2_{k_1}\ldots w^2_{k_6}\ 
\prod_{t=1}^{d-1} |T_t^{\varphi}| \ssj{k_1}{k_2}{k_3}{k_6}{k_4}{k_5} = Z(\M)\cdot f(S^2_{(4,6,4)},j)\]
\\
since

\[ \prod_{s=f+1}^{b-6}w^2_{\varphi(E_s)}\ w^2_{k_1}\ldots w^2_{k_6}=
\prod_{s=f+1}^b w^2_{\varphi(E_s)} \]

\[ \prod_{t=1}^{d-1} |T_t^{\varphi}| \ssj{k_1}{k_2}{k_3}{k_6}{k_4}{k_5}= 
 \prod_{t=1}^d |T_t^{\varphi}|  \]
\\
\\
It is easy to see that the last result can be generalized to an arbitrary $(N_0, N_1, N_2)$ spherical boundary (this can be done by gluing to the (4,6,4) boundary of $\M'$ a hypercylinder $S^2\times [0,1]$ having the 2 boundaries triangulated as $S^2_{(4,6,4)}$ and $S^2_{(N_0,N_1,N_2)}$, respectively). Therefore, the general result is:

\begin{equation}
Z(\M') = Z(\M)\cdot f(S^2_{(N_0,N_1,N_2)},j)
\end{equation}

\subsection{Connected sum}

Let $\M$ and $\N$ be two arbitrary 3-manifolds. Then the partition function for the connected sum $\M \# \N$ is 

\begin{equation}
Z(\M \# \N)=\frac{Z(\M)\ Z(\N)}{Z(S^3)}
\label{m_and_n}
\end{equation}
\\
{\bf Proof:}

By making an $S^2$ boundary (i.e. cutting out a 3-ball) in both manifolds and then gluing them together on the common boundary we get:

\[ Z(\M \# \N) = \sum_i Z(\M;S^2,i)\ Z(\N;S^2,i)= \]
\[ =Z(\M)\  Z(\N) \sum_i f^2(S^2,i) = \frac{Z(\M)\ Z(\N)}{Z(S^3)} \]
\\
{\bf Observation:}

This result is similar to that from 2 dimensions, with $Z(S^3)$ replaced by $Z(S^2)$. In 2 dimensions, this follows from

\[\chi(\M \# \N)=\chi(\M) + \chi(\N) - \chi(S^2)\]
\\
and from the fact that the partition function is $Z(\M)\sim e^{\chi(\M)}$.
Eq.~(\ref{m_and_n}) is also identical to a result obtained by Witten \cite{witten}.

\subsection{Adding/removing a 3-handle $S^2\times S^1$}

Adding a handle is done simply by cutting out two $S^2$ boundaries and then gluing on them the hypercylinder $S^2\times [0,1]$. From (\ref{zs32}), (\ref{s2factor}) and (\ref{idf}) we have

\[ Z(\M \cup (S^2\times S^1))=Z(\M)\ Z(S^3) \sum_{j, k \in I} f^2(S^2_{(4,6,4)},j)\ f^2(S^2_{(4,6,4)},k)= \]

\[ =Z(\M)\cdot c\ w^2\]

Alternatively, the last equation follows directly from (\ref{m_and_n}) if we view the resulting manifold as a connected sum $\M \# (S^2 \times S^1)$.

Removing the handle is the reverse operation and therefore amounts to multiplying the partition function by $c^{-1}\ w^{-2}$.

\subsection{Making a $T_g$ boundary}

We would like to obtain a result similar to that of Section \ref{s2bound} for an arbitrary $T_g$ boundary. However, this factorization does not hold in general, but only for a special case, namely  when the $T_g$ boundary can be contained in a 3-ball ${\cal B}^3$.

With this proviso, we can decompose $\M'$ by writing (See Fig. \ref{tg2s2}):

\begin{figure}
\epsffile{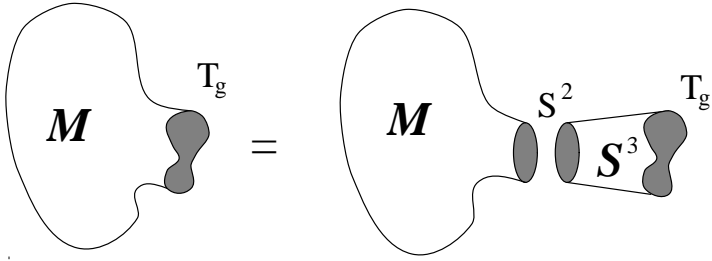}
\caption{Factorization of a $T_g$ boundary.}
\label{tg2s2}
\end{figure}

\begin{equation}
Z(\M; T_g,i)=\sum_{\scriptsize \begin{array}{c} colourings\\ j \in adm(S^2) \end{array}}Z(\M; S^2,j)\cdot Z(S^3;\ S^2,j,\ T_g,i)
\end{equation}

Each term of the sum is a manifold with an $S^2$ boundary, so we can apply a previous lemma and factor out the boundaries contribution

\[ Z(\M; T_g,i)=Z(\M)\ Z(S^3; T_g,i)\ \sum_j f^2(S^2,j)= \]
\[ =\frac{Z(\M)\ Z(S^3; T_g,i)}{Z(S^3)}=Z(\M)\ g(T_g,i) \]
where
\[ g(T_g,i) \equiv \frac{Z(S^3;T_g,i)}{Z(S^3)} \]
\\
is obviously a function only of the coloring $i$ of $T_g$.\\
\\
{\bf Observation:}

In the case of a g-genus handlebody (i.e. a 3-sphere with a $T_g$ boundary) we cannot infer a relation similar to eq.~(\ref{idf}) for the function $g$. Although gluing 2 handlebodies on the common $T_g$ boundary yields a 3-sphere $S^3$, the gluing is done with a twist, since we identify the coordinates $(x,y)$ on one of them with $(y',x')$ on the other.

\sect{Conclusions}

In this paper we propose a general scheme for calculating the Turaev-Viro invariant $Z(\M)$ for a 3-manifold $\M$. This consist of finding a list of simple manifolds with known $Z(\M)$ and then performing 'surgeries' on these, together with the induced modification of such operations on the invariant. This framework is quite general and can be applied to other invariants. Of course, we need a proof of the completness of this scheme, i.e. that any 3-manifold can be constructed in this way.

Analyzing the making of a boundary, we proved that for $S^2$ boundaries the invariant $Z(\M;S^2,\ldots)$ factors out the contribution of each boundary, the remaining part being the Turaev-Viro invariant $Z(\M)$ of a closed manifold.

This factorization property enables us to prove by direct calculation the formula for the connected sum of two manifolds $Z(\M \# \N)$. Other surgeries are the attaching/removing of a 3-handle $S^2\times S^1$ and the making of a $T_g$ boundary.

However, for a general $T_g$ boundary the factorization theorem is not true in general, although it holds in a particular case. Making a $T_g$ boundary in an arbitrary manifold $\M$ depends on the topology of $\M$ and also on the existence of other $T_g$ boundaries. But this will be investigated somewhere else.\\
\\
{\Large\bf Acknowledgements} \\
\\
This work has been kindly supported by Cambridge Overseas Trust, The Ra\c tiu Foundation and ORS. The author wishes to thank Dr.~Ruth Williams for helpful discussions and permanent encouragement.

\newpage
\appendix

{\Large\bf Appendix 1} \\

Calculation of the partition function $Z(S^3;S^2,\ldots,\ S^2)$ for $S^2$ boundaries. \\

a) One boundary $Z(S^3;\ S^2,j)$\\
   
In this case we have (see Fig. \ref{s3s2}):

\[ Z(S^3;\ S^2,j)=w^{-6}\ w_{j_1} \ldots w_{j_6} \sum_{k_1...k_4}w^2_{k_1}\ldots w^2_{k_4} \ssj{k_1}{k_2}{j_1}{j_5}{j_4}{k_4} \cdot \]
\[ \ssj{k_1}{k_3}{j_3}{j_6}{j_4}{k_4} \ssj{k_2}{k_3}{j_2}{j_6}{j_5}{k_4} \ssj{j_1}{j_2}{j_3}{k_3}{k_1}{k_2} \]

By summing over $j_3$ in the last three $6j$ symbols we get:

\[ \sum_{j_3} w^2_{j_3}\ \ssj{k_1}{k_3}{j_3}{j_6}{j_4}{k_4} \ssj{k_2}{k_3}{j_2}{j_6}{j_5}{k_4} \ssj{j_1}{j_2}{j_3}{k_3}{k_1}{k_2}= \]

\[ =\ssj{j_1}{j_4}{j_5}{k_4}{k_2}{k_1} \ssj{j_1}{j_4}{j_5}{j_6}{j_2}{j_3}= 
\ssj{k_1}{k_2}{j_1}{j_5}{j_4}{k_4} \ssj{j_1}{j_2}{j_3}{j_6}{j_4}{j_5} \]
\\
the last equality resulting from the symmetry properties of the $6j$ symbols.
Therefore we obtain

\[Z(S^3;\ S^2,j)=w^{-6}\ w_{j_1} \ldots w_{j_6} \ssj{j_1}{j_2}{j_3}{j_6}{j_4}{j_5} \sum_{k_1,k_2,k_4}w^2_{k_1}\ w^2_{k_2}\ w^2_{k_4} \ssj{k_1}{k_2}{j_1}{j_5}{j_4}{k_4}^2 =\]

\[ =w^{-6}\ w_{j_1} \ldots w_{j_6} \ssj{j_1}{j_2}{j_3}{j_6}{j_4}{j_5}\ \sum_{k_1,k_2} \frac{w^2_{k_1}\ w^2_{k_2}}{w^2_{j_1}} = 
w^{-4}\ w_{j_1} \ldots w_{j_6} \ssj{j_1}{j_2}{j_3}{j_6}{j_4}{j_5}\]
\\

b) Two boundaries $Z(S^3;\ S^2,j,\ S^2,k)$\\

\begin{figure}
\epsffile{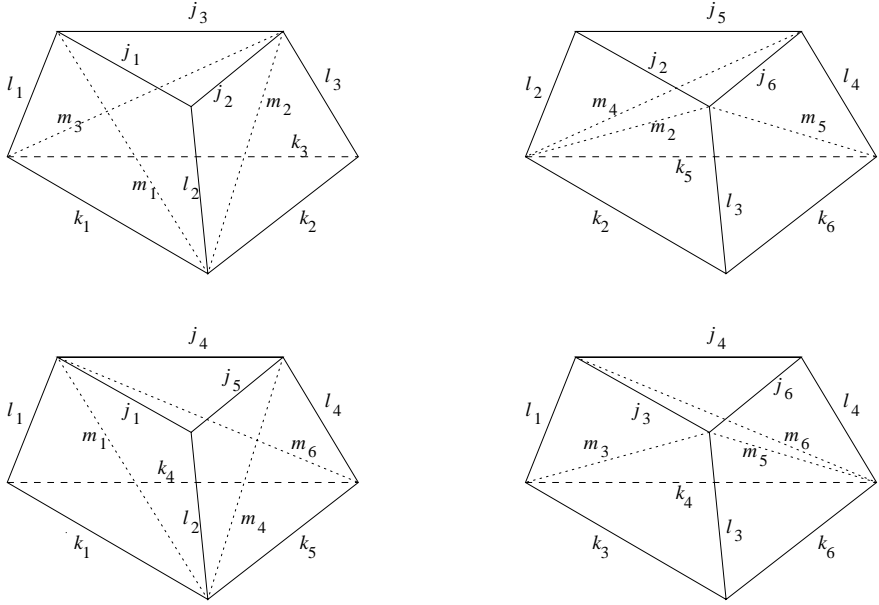}
\caption{Decomposition of $S^3$ into tetrahedra.}
\label{tetra4}
\end{figure}

In this case we have a 3 sphere $S^3$ with two $S^2$ boundaries. The triangulation is given by $N_0=8$ vertices, $N_1=22$ edges, $N_2=28$ triangles and $N_3=12$ tetrahedra (see Fig. \ref{tetra1}). The boundaries have the edges ${j_1,...,j_6}$ and ${k_1,...,k_6}$. Intuitively, this can be viewed as a sphere $S^2$ inside another sphere. Each vertex of one boundary is joined with the corresponding one of the other boundary by the edges ${l_1,...,l_4}$. Thus, to each face of the boundary corresponds a prism with one basis on the $S^2$ boundary and the other one on the remaning $S^2$ boundary. Each such prism is further decomposed into three tetrahedra, the new edges being ${m_1,...,m_6}$ (see Fig. \ref{tetra4}) and we get

\[Z(S^3;S^2,j,\ S^2,k)=w^{-8} \sum_{\scriptsize \begin{array}{c}l_1...l_4 \\ m_1...m_6 \end{array}} w_{j_1}\ldots w_{j_6}\ w_{k_1}\ldots w_{k_6}\ w^2_{l_1}\ldots w^2_{l_4}\ w^2_{m_1}\ldots w^2_{m_6} \cdot \]
\[ \ssj{j_1}{j_2}{j_3}{m_2}{m_1}{l_2} \ssj{k_1}{k_2}{k_3}{l_3}{m_3}{m_2} \ssj{l_1}{m_1}{k_1}{m_2}{m_3}{j_3} \ssj{j_2}{j_5}{j_6}{m_4}{m_2}{l_2} \cdot \]
 
\[ \ssj{k_2}{k_5}{k_6}{m_5}{l_3}{m_2} \ssj{m_2}{m_4}{j_6}{l_4}{m_5}{k_5} \ssj{j_1}{j_4}{j_5}{m_4}{l_2}{m_1} \ssj{k_1}{k_4}{k_5}{m_6}{m_1}{l_1} \cdot \]

\begin{equation} 
\ssj{m_1}{m_6}{k_5}{l_4}{m_4}{j_4} \ssj{k_3}{k_4}{k_6}{m_5}{l_3}{m_3}
\ssj{j_3}{j_4}{j_6}{l_4}{m_5}{m_6} \ssj{l_1}{j_3}{m_3}{m_5}{k_4}{m_6}
\label{big12j}
\end{equation}

Now we can apply four times the Biedenharn-Elliot identity:

\[\sum_{l_1}w^2_{l_1}\ \ssj{l_1}{m_1}{k_1}{m_2}{m_3}{j_3} \ssj{k_1}{k_4}{k_5}{m_6}{m_1}{l_1} \ssj{l_1}{j_3}{m_3}{m_5}{k_4}{m_6} =
\ssj{m_2}{m_5}{k_5}{m_6}{m_1}{j_3} \ssj{m_2}{m_5}{k_5}{k_4}{k_1}{m_3} \]

\[\sum_{l_2}w^2_{l_2}\ \ssj{j_1}{j_2}{j_3}{m_2}{m_1}{l_2} \ssj{j_2}{j_5}{j_6}{m_4}{m_3}{l_2} \ssj{j_1}{j_4}{j_5}{m_4}{l_2}{m_1} =
\ssj{j_1}{j_2}{j_3}{j_6}{j_4}{j_5} \ssj{m_1}{m_2}{j_3}{j_6}{j_4}{m_4} \]

\[\sum_{l_3}w^2_{l_3}\ \ssj{k_1}{k_2}{k_3}{l_3}{m_3}{m_2} \ssj{k_2}{k_5}{k_6}{m_5}{l_3}{m_2} \ssj{k_3}{k_4}{k_6}{m_5}{l_3}{m_3} =
\ssj{k_1}{k_4}{k_5}{m_5}{m_2}{m_3} \ssj{k_1}{k_4}{k_5}{k_6}{k_2}{k_3} \]

\[\sum_{l_4}w^2_{l_4}\ \ssj{m_2}{m_4}{j_6}{l_4}{m_5}{k_5} \ssj{m_1}{m_6}{k_5}{l_4}{m_4}{j_4} \ssj{j_3}{j_4}{j_6}{l_4}{m_5}{m_6} =
\ssj{m_1}{m_2}{j_3}{m_5}{m_6}{k_5} \ssj{m_1}{m_2}{j_3}{j_6}{j_4}{m_4} \]

From the symmetry of the $6j$ symbols

\[ \ssj{m_2}{m_5}{k_5}{m_6}{m_1}{j_3}=\ssj{m_1}{m_2}{j_3}{m_5}{m_6}{k_5}=\ssj{m_1}{j_3}{m_2}{m_5}{k_5}{m_6} \]

\[ \ssj{m_2}{m_5}{k_5}{k_4}{k_1}{m_3}=\ssj{k_1}{k_4}{k_5}{m_5}{m_2}{m_3}=\ssj{m_2}{k_1}{m_3}{k_4}{m_5}{k_5} \]

\[ \ssj{m_1}{m_2}{j_3}{j_6}{j_4}{m_4}= \ssj{m_1}{j_4}{m_4}{j_6}{m_2}{j_3} \]

Thus, we obtain

\[ Z=w^{-8}\ w_{j_1}\ldots w_{j_6}\ w_{k_1}\ldots w_{k_6} \ssj{j_1}{j_2}{j_3}{j_6}{j_4}{j_5} \ssj{k_1}{k_4}{k_5}{k_6}{k_2}{k_3} \cdot \]
\[ \sum_{m_1...m_6}\ w^2_{m_1}\ldots w^2_{m_6}\ \ssj{m_2}{k_1}{m_3}{k_4}{m_5}{k_5}^2 
\ssj{m_1}{j_4}{m_4}{j_6}{m_2}{j_3}^2 \ssj{m_1}{j_3}{m_2}{m_5}{k_5}{m_6}^2 \]

The last sum can be written as (after applying the orthogonality relations and summing over $m_3,\ m_4$):

\[ \sum_{\scriptsize \begin{array}{c}m_1,m_2 \\ m_5,m_6 \end{array}} \frac{w^2_{m_1} w^2_{m_2} w^2_{m_5} w^2_{m_6}}{w^2_{k_5} w^2_{j_3}} \ssj{m_1}{j_3}{m_2}{m_5}{k_5}{m_6}^2= \frac{1}{w^2_{k_5}w^2_{j_3}} \sum_{\scriptsize \begin{array}{c} m_1,m_5, \\ m_6 \in adm \end{array}}w^2_{m_1}w^2_{m_5}= \]
\[ =\frac{1}{w^2_{k_5}w^2_{j_3}}\ \sum_{m_5,m_6} c\ w^2_{m_6}w^2_{k_5}w^2_{m_5} \delta (m_5, m_6, j_3)= c\ w^2 \]
\\
and we used the relations:

\[ \sum_{m_1}w^2_{m_1} \delta(m_1,m_6,k_5)= c\ w^2_{k_5}w^2_{m_6} \]
\[ \sum_{m_5,m_6}\ w^2_{m_5}w^2_{m_6} \delta (m_5, m_6, j_3)= w^2\ w^2_{j_3} \]

Therefore, the partition function is

\begin{equation}
Z(S^3;S^2,j,\ S^2,k)= c\ w^{-6}\ w_{j_1}\ldots w_{j_6}\ w_{k_1}\ldots w_{k_6} \ssj{j_1}{j_2}{j_3}{j_6}{j_4}{j_5} \ssj{k_1}{k_4}{k_5}{k_6}{k_2}{k_3}
\label{zcil}
\end{equation}

\newpage
\appendix

{\Large\bf Appendix 2} \\

The partition function $Z(S^3;\ S^2_{(N_0,N_1,N_2)},j)$ of the 3-ball ${\cal B}^3$ for different triangulations of the $S^2$ boundary
\\

a) (3,3,2)\\

In this case we have (see Fig. \ref{s3s2}):\\

$a=4$, $e=3$, $b=6$, $f=3$, $d=2$

\[ Z=w^{-5}\ w_{j_1}w_{j_2}w_{j_3} \sum_{k_1,k_2,k_3} w^2_{k_1}w^2_{k_2}w^2_{k_2} \ssj{j_1}{j_2}{j_3}{k_3}{k_2}{k_1}^2= \]

\[ =w^{-5}\ w_{j_1}w_{j_2}w_{j_3} \sum_{k_2,k_3} \frac{w^2_{k_2}w^2_{k_3}}{w^2_{j_3}}=w^{-3}\ w_{j_1}w_{j_2}w_{j_3} \]
\\

b) (5,9,6)\\

$a=6$, $e=5$, $b=14$, $f=9$, $d=6$

\[ Z=w^{-7}\ w_{j_1}\ldots w_{j_9}\ \sum_{k_1 \ldots k_5} w^2_{k_1}\ldots w^2_{k_5} \ssj{k_1}{k_2}{j_3}{j_4}{j_1}{k_4} \ssj{j_2}{j_3}{j_5}{k_2}{k_3}{k_1} \cdot\]

\[ \ssj{j_1}{j_2}{j_6}{k_3}{k_4}{k_1} \ssj{j_4}{j_7}{j_8}{k_5}{k_2}{k_4} \ssj{j_5}{j_8}{j_9}{k_5}{k_3}{k_2} \ssj{j_6}{j_7}{j_9}{k_5}{k_3}{k_4}\]

Applying the Biedenharn-Elliot identity in the forms

\[ \sum_{k_1} w^2_{k_1} \ssj{k_1}{k_2}{j_3}{j_4}{j_1}{k_4} \ssj{j_2}{j_3}{j_5}{k_2}{k_3}{k_1}
\ssj{j_1}{j_2}{j_6}{k_3}{k_4}{k_1} = \ssj{j_4}{j_5}{j_6}{j_2}{j_1}{j_3} \ssj{j_3}{j_5}{j_6}{k_3}{k_4}{k_2} \]

\[ \sum_{k_5}w^2_{k_5} \ssj{j_4}{j_7}{j_8}{k_5}{k_2}{k_4} \ssj{j_5}{j_8}{j_9}{k_5}{k_3}{k_2}
\ssj{j_6}{j_7}{j_9}{k_5}{k_3}{k_4} = \ssj{j_4}{j_5}{j_6}{j_9}{j_7}{j_8} \ssj{j_3}{j_5}{j_6}{k_3}{k_4}{k_2} \]
\\
we get

\[ Z=w^{-7}\ w_{j_1}\ldots w_{j_9} \ssj{j_4}{j_5}{j_6}{j_2}{j_1}{j_3} \ssj{j_4}{j_5}{j_6}{j_9}{j_7}{j_8} \sum_{k_2,k_3,k_4} w^2_{k_2}w^2_{k_3}w^2_{k_4} \ssj{j_4}{j_5}{j_6}{k_3}{k_4}{k_2}^2= \]

\[ =w^{-5}\ w_{j_1}\ldots w_{j_9} \ssj{j_4}{j_5}{j_6}{j_2}{j_1}{j_3} \ssj{j_4}{j_5}{j_6}{j_9}{j_7}{j_8} \]


\begin{thebibliography}{}

\bibitem{ish1} C. Isham, Structural Issues in Quantum Gravity, gr-qc/9510063

\bibitem{ish2} C. Isham, Prima Facie Questions in Quantum Gravity, gr-qc/9310031

\bibitem{jb} J. W. Barrett, J. Math. Phys. {\bf 36} (1995), 6161

\bibitem{pr} G. Ponzano and T. Regge, "Semiclassical limit of Racah coefficients", in {\it Spectroscopic and Group Theoretical Methods in Physics}, F.Bloch, ed. (North-Holland, Amsterdam, 1968), pp. 1-58

\bibitem{tv} V. G. Turaev and O. Y. Viro, Topology, {\bf 31} (1992), 865

\bibitem{witten} E. Witten, Comm. Math. Phys. {\bf 117} (1989), 351

\bibitem{kms} M. Karowski, W. M\"uller and R. Schrader, J. Phys. {\bf A25} (1992), 4847

\end{thebibliography}
\end{document}